\documentclass[12pt]{article}
\usepackage{subfiles}
\usepackage{amsmath,amssymb}
\usepackage{physics}
\usepackage{enumerate}
\usepackage{siunitx}
\usepackage{tikz}
\usepackage{graphicx,caption}
\usepackage{slashed}
\usepackage{hyperref}
\usepackage{float}
\usepackage{subcaption}
\usepackage[margin=20truemm]{geometry}
\usepackage[style=phys,biblabel=brackets,chaptertitle=false,pageranges=false,eprint]{biblatex}

\numberwithin{equation}{section}

\begin{document}

\begin{flushright}
\hfill{KUNS-3036}
\end{flushright}
\begin{center}
\vspace{2ex}
{\Large \textbf{Generalized Collective Coordinate Quantization of Solitons with Topological Terms
}}

\vspace*{5mm}
\textsc{Taichi Tsukamoto}$^{a}$\footnote{e-mail:
 \texttt{taichi.t@gauge.scphys.kyoto-u.ac.jp}}
\vspace*{4mm}

\hspace{-0.5cm}
\textit{{$^a$
Department of Physics, Kyoto University, Kyoto 606-8502, Japan
}}\\
\end{center}

\vspace*{.5cm}

\begin{abstract}
    We show that the $\theta^+$ pentaquark does not exist in the Skyrme model. For the solitons of the theory with topological terms, the standard collective coordinate quantization does not construct the proper low energy effective theory. In the presence of topological terms, zero modes are classified into three groups: dynamical zero modes with constant velocity, cyclotron zero modes in a circular orbit, and static constraint zero modes. According to the mode expansion, the topological term contributes to the moduli metric of the dynamical zero modes. In addition, constraint zero modes do not have the kinetic term and produce constraints that strongly restrict the spectrum instead. In this paper, we propose a generalized collective coordinate quantization method and apply it to the $SU(3)$ Skyrmion with the Wess-Zumino-Witten term. We find five constraints. These reproduce the results of \cite{Cherman:2005hy}, eliminating all the $SU(3)$ multiplets except the nucleon octet and $\Delta$ baryon decuplet from the baryon spectrum.
\end{abstract}

\newpage

\section{Introduction}
One of the basic methods to analyze the dynamics of solitons is the collective coordinate quantization (CCQ) \footnote{See, e.g. \cite{Manton:2004tk} for a review.}. Generally, solitons are parametrized by several collective coordinates, and the set of solitons with the same energy is called the moduli space. Collective coordinates are related to the zero modes of the fluctuations around the soliton and are important in the low energy dynamics. In the CCQ method, the low energy effective theory is constructed by introducing the time-dependent collective coordinates to the soliton and putting them into the action. Solitons are described as non-relativistic particles slowly moving in the moduli space in this approach. 

However, the analysis based on the CCQ might not be reliable for solitons with topological terms, especially the chiral soliton models in quantum chromodynamics (QCD). One example is the chiral soliton model with $\rho$ and $\omega$ mesons with the Wess-Zumino-Witten (WZW) term in two flavors shown in \cite{Meissner:1987ge}. Solitons are interpreted as baryons and have $SU(2)$ collective coordinates that correspond to the spin/isospin degrees of freedom \cite{Adkins:1983ya}. In a classical soliton configuration, the space components of $\omega$ meson are in pure gauge. When promoting the time-dependent collective coordinates, however, the space components of the $\omega$ meson field strength become non-zero valued proportional to the angular velocity since the topological charge behaves as a $U(1)$ charge of the $\omega$ meson. Therefore, the momentum of inertia has the contribution from $\omega$. The same phenomenon can be seen in holographic QCD \cite{Panico:2008it}.

Another example is the $SU(3)$ Skyrmion with the WZW term. According to the naive CCQ, a constraint arises from the WZW term and indicates that the baryon state with baryon number $N_B=1$ must have the particle with hypercharge $Y=N_c/3$ in the representation \cite{Guadagnini:1983uv,Chemtob:1985ar,Diakonov:1997mm}. In $N_c=3$ case, the allowed multiplets are $8$, $10$, $\overline{10}$, $\cdots$. $8$ and $10$ include nucleons and $\Delta$ baryons, respectively, and $\overline{10}$ predicts exotic hadrons such as $\theta^+$ pentaquark. $\theta^+$ is the particle with spin $\frac{1}{2}$, isospin $0$, and strangeness $+1$. Although many experiments were held to search $\theta^+$, only a few reported the existence of this particle \footnote{For the experiments and their results, see \cite{Praszalowicz:2024zsy} or \cite{Amaryan:2025muw}, for example}.
On the theoretical side, the existence of $\theta^+$ depends on the models or methods used to analyze the spectrum of the baryon. In the bound state approach (BSA) for the Skyrme model, no bound states or resonances are found that can be interpreted as $\theta^+$ \cite{Callan:1985hy,Callan:1987xt}. Cherman, Cohen, Dulaney, and Lynch pointed out that some zero modes of the Skyrmion are not dynamical \cite{Cherman:2005hy}. Distinguishing dynamical and non-dynamical zero modes, they found the upper bound on the hypercharge of the baryons that prohibits all the exotic baryons including $\theta^+$ pentaquark. Opposing them, Walliser and Weigel introduced the rotation-vibration approach (RVA) and calculated the mass and the width of $\theta^+$ \cite{Walliser:2005pi}.

In this paper, we improve the CCQ method in the presence of the topological term by solving the EOM of the fluctuation around the soliton. It turns out that zero modes are classified into three groups due to the form of the solution of the EOM. In addition, the zero mode action and the moduli metric constructed from the mode expansion do not coincide with those derived from the naive CCQ. Our approach extends \cite{Cherman:2005hy} to be able to quantize properly.

The organization of the paper is as follows. First, we discuss the spectrum of the fluctuation around the soliton in section \ref{sec:fluctuation}, and see that there are several types of zero modes. In section \ref{sec:CCQ}, we construct the zero mode action, which describes the low energy dynamics of solitons correctly, by the mode expansion. The generalization of the CCQ method is also introduced in this section. This is applied to the $SU(3)$ Skyrme-WZW model in section \ref{sec:Skyrmion}, and we show that $\theta^+$ pentaquark is eliminated from the spectrum. We also compare our method with other approaches and discuss the relation between them in this section. Section \ref{sec:conclusion} is devoted to the conclusion and discussion of possible future directions.
\section{The spectrum of the fluctuation}
\label{sec:fluctuation}
\subsection{The quadratic Lagrangian}
\label{subsec:fluctuation}
The basic idea of the collective coordinate quantization is the quantum theory of fluctuations around the soliton solution. Consider an action of bosonic variables
\begin{equation}
\label{eq:theory-original}
    S[\phi] = \int\dd t L(\phi(t),\dot{\phi}(t)), \quad\phi = \qty{\phi^x}.
\end{equation}
Here, $x$ is the index of $\phi$.
In quantum field theory (QFT), $x$ denotes the spatial coordinates and indices of internal symmetries, for example.
The equation of motion (EOM) is 
\begin{equation}
    \dv{t}\fdv{L}{\dot{\phi}}(\phi,\dot{\phi}) -\fdv{L}{\phi}(\phi,\dot{\phi}) = 0.
\end{equation}
Assume that this has the static solution $\phi = \phi_\mathrm{sol}$, $\dot{\phi}_\mathrm{sol} = 0$. The static EOM is 
\begin{equation}
\label{eq:EOM-static}
    \dv{t}\fdv{L}{\dot{\phi}}(\phi,0) = \fdv{L}{\phi}(\phi,0) = 0.
\end{equation}
Note that the solution of the static EOM \eqref{eq:EOM-static} with given energy is not unique in general and has several parameters. In QFT, these parameters are collective coordinates and the space of collective coordinates is called the moduli space of solitons. For example, the collective coordinate and the moduli space of a kink on a $1+1$ dimensional spacetime is the position of the kink $\mathbb{R}$. 

Let the fluctuation around the static solution be $\phi_\mathrm{sol}$ as $q(t):=\phi(t)-\phi_\mathrm{sol}$. Substituting into the action, we have 
\begin{equation}
\label{eq:action-fluctuation}
    S[\phi_\mathrm{sol}+q]-S[\phi_\mathrm{sol}] \simeq \int\dd t\qty[\frac{1}{2}\qty(\dot{q}^TM\dot{q}-q^TKq+q(A+B)\dot{q})+P_0^T\dot{q}],
\end{equation}
in the leading order of $q$. Here, $M$, $K$, $A$, $B$, and $P_0$ are the variations of the Lagrangian
\begin{equation}
    M_{xy}:=\frac{\var^2L}{\var\dot{\phi}^x\var\dot{\phi}^y}(\phi_\mathrm{sol},0),\,K_{xy}:=-\frac{\var^2L}{\var\phi^x\var\phi^y}(\phi_\mathrm{sol},0),
\end{equation}
\begin{equation}
    \,\,A_{xy} := \frac{\var^2L}{\var\phi^x\var\dot{\phi}^y}(\phi_\mathrm{sol},0)+\frac{\var^2L}{\var\dot{\phi}^x\var\phi^y}(\phi_\mathrm{sol},0),\,B_{xy} := \frac{\var^2L}{\var\phi^x\var\dot{\phi}^y}(\phi_\mathrm{sol},0)-\frac{\var^2L}{\var\dot{\phi}^x\var\phi^y}(\phi_\mathrm{sol},0),
\end{equation}
\begin{equation}
    (P_0)_x:={\fdv{L}{\dot{\phi}^x}}(\phi_\mathrm{sol},0),
\end{equation}
respectively. By definition, $M$, $A$, and $K$ are symmetric and $B$ is an antisymmetric operator. To guarantee the stability of $\phi_\mathrm{sol}$ and the non-existence of gauge degrees of freedom or constraints, we assume that $M$ is positive definite and $K$ is positive semidefinite. In the standard soliton models such as kinks \cite{Gervais:1974dc} or $SU(2)$ Skyrmions \cite{Adkins:1983ya}, there are only $M$ and $K$ terms, which are the kinetic term and harmonic potential term of $q$, respectively. $A$, $B$, and $P_0$ terms come from the term with the first order of time derivatives in the original theory \eqref{eq:theory-original}. These terms behave as the constant magnetic field in the $q$ target space. In QFT, they appear in the presence of metric-independent topological terms such as $\theta$ terms, WZW terms, and Chern-Simons terms. 

The EOM from \eqref{eq:action-fluctuation} is 
\begin{equation}
\label{eq:EOM_q}
    M\ddot{q}-B\dot{q}+Kq = 0.
\end{equation}
Note that $P_0$ and the symmetric $A$ do not contribute to the EOM, since $P_0^T\dot{q}$ and $q^TA\dot{q}$ are actually total derivative terms. For simplicity, we will neglect these terms.

The EOM \eqref{eq:EOM_q} is a linear differential equation with constant coefficients. It is convenient to substitute $q=e^{-i\omega t}\xi$ into \eqref{eq:EOM_q} and we have
\begin{equation}
\label{eq:EOM-eigenvector}
    (-\omega^2M+i\omega B+K)\xi = 0.
\end{equation}
For this equation to have a non-trivial ``eigenvector'' $\xi$, $\omega$ must be the root of the following characteristic equation 
\begin{equation}
\label{eq:EOM-det}
    \Phi(\omega) := \det(-\omega^2M+i\omega B+K) = 0.
\end{equation}
When $B=0$, this is merely an eigenvalue problem of $M^{-1}K$. 
We will mention some properties of \eqref{eq:EOM-eigenvector} and \eqref{eq:EOM-det}.
First, $\Phi(\omega)$ is an even polynomial of $\omega$. It is shown by
\begin{equation}
\begin{split}
    \Phi(\omega) & = \det(-\omega^2M+i\omega B+K) = \det((-\omega^2M+i\omega B+K)^T)\\
    & = \det(-\omega^2M-i\omega B+K) = \Phi(-\omega).
\end{split}
\end{equation}

In addition, the roots of $\Phi(\omega)$ are real. Let $(\omega,\xi)$ be the solution of \eqref{eq:EOM-eigenvector} and \eqref{eq:EOM-det}. Taking the inner product of $\xi^\dagger$ and \eqref{eq:EOM-eigenvector},
\begin{equation}
    -\omega^2\xi^\dagger M\xi+i\omega\xi^\dagger B\xi+\xi^\dagger K\xi = 0.
\end{equation}
The discriminant of this equation is $-(\xi^\dagger B\xi)^2+4(\xi^\dagger M\xi)(\xi^\dagger K\xi)$. Since $M$ and $K$ are positive semidefinite and $B$ is antisymmetric, $\xi^\dagger M\xi$ and $\xi^\dagger K\xi$ are non-negative and $\xi^\dagger B\xi$ is pure imaginary. Thus, the discriminant is always non-negative and $\omega$ is real. Taking the complex conjugate of \eqref{eq:EOM-eigenvector}, we have
\begin{equation}
    (-\omega^2M-i\omega B+K)\xi^* = 0.
\end{equation}
Therefore, $(-\omega,\xi^*)$ also satisfies \eqref{eq:EOM-eigenvector} and \eqref{eq:EOM-det}. 

Under the mode expansion, zero modes $\omega=0$ play the central role in CCQ. Zero modes are characterized by $Kq=0\Leftrightarrow q=\ker K$. The constant fluctuations along the $\ker K$ direction do not change the energy of $\phi_\mathrm{sol}$. Thus, zero modes correspond to the infinitesimal transformation of collective coordinates of $\phi_\mathrm{sol}$, and $\dim\ker K$ is equal to the dimension of the moduli space. We denote the indices of the collective coordinates and the basis of $\ker K$ by $a, b, \cdots,$ and $\qty{X^a}$, $\qty{\xi_a}$, respectively. We can take the basis $\qty{\xi_a}$ as
\begin{equation}
    \xi_a=\pdv{\phi_\mathrm{sol}}{X^a},
\end{equation}
for example.
\subsection{Contradiction of the naive collective coordinate quantization}
Now we discuss the naive CCQ to show that this method has a critical contradiction.
In a naive CCQ, the effective action is acquired by introducing the time-dependent collective coordinates $\phi =\phi_\mathrm{sol}(X^a(t))$ and putting them into the original action. 
This corresponds to assuming $q(t)$ takes the form
\begin{equation}
\label{eq:naive-CCQ_q}
    q(t)=\sum_aX^a(t)\pdv{\phi_\mathrm{sol}}{X^a}=\sum_aX^a(t)\xi_a.
\end{equation}
Putting it into \eqref{eq:action-fluctuation}, we can evaluate the naive CCQ action $S_\mathrm{naive}[X^a]:=S[\phi_\mathrm{sol}(X^a(t))]$ as
\begin{equation}
\label{eq:eff-action-naive}
    S_\mathrm{naive}[X^a]\simeq\int\dd t\frac{1}{2}\sum_{a,b}\qty(M_{ab}\dot{X}^a\dot{X}^b+B_{ab}X^a\dot{X}^b),
\end{equation}
where $M_{ab}:=\xi_a^TM\xi_b,B_{ab}:=\xi_a^TB\xi_b$. 
The EOM of $S_\mathrm{naive}$ is 
\begin{equation}
\label{eq:EOM-naiveCCQ}
    M_{ab}\ddot{X}^b-B_{ab}\dot{X}^b = 0.
\end{equation}
This is an EOM of a charged particle with a curved background metric $M_{ab}$ and a constant magnetic field $B_{ab}$.

However, substituting \eqref{eq:naive-CCQ_q} into \eqref{eq:EOM_q}, we have
\begin{equation}
\label{eq:EOM-naive-zero}
    \sum_a\ddot{X}^a M\xi_a-\dot{X}^a B\xi_a = 0.
\end{equation}
The only solutions of \eqref{eq:EOM-naive-zero} are $\dot{X}^a\equiv0$, since $M\xi_a$ and $B\xi_a$ are linearly independent in general \cite{Cherman:2005hy}. Thus, the time-dependent part of the solution of \eqref{eq:EOM-naiveCCQ} cannot be the solution of \eqref{eq:EOM-naive-zero}. In addition, $\dot{X}^a\equiv0$ claims that there are no dynamics in zero modes, which contradicts the dynamics that solitons can actually have. For example, solitons have the freedom of center of mass coordinates when the theory has Poincar\'{e} symmetry. Using the naive CCQ, \eqref{eq:EOM-naive-zero} implies that solitons cannot move at finite velocity when $B\neq0$. However, one can easily construct the exact soliton solution moving with constant velocity through a Lorentz transformation. 

Another disadvantage of the naive CCQ appears when we consider other $\omega\neq0$ modes along with zero modes.
One might propose the following ansatz
\begin{equation}
\label{eq:RVA-ansatz}
    q = \sum_aX^a(t)\xi_a+q', \xi_a^TMq' = 0,
\end{equation}
and regard $q'$ as $\omega\neq0$ modes.
The action \eqref{eq:action-fluctuation} would be
\begin{equation}
\begin{split}
\label{eq:eff-action-naive-with-nonzero}
    S & = S_\mathrm{naive}+\int\dd t\frac{1}{2}\qty(\dot{q}'^TM\dot{q}'+q'^TB\dot{q}'-q'^TKq')\\
    & +\int\dd t\frac{1}{2}\qty(q'^TB\xi_a\dot{X}^a-\dot{q}'^TB\xi_aX^a).
\end{split}
\end{equation}
The first line is $S_\mathrm{naive}$ defined in \eqref{eq:eff-action-naive} and the action of $q'$, respectively. The second line, on the other hand, seems to be the ``interaction'' term between $q'$ and zero modes. Generally, this ``interaction'' term is not small compared to other terms and we cannot treat it perturbatively.

Therefore, the naive CCQ $\phi_\mathrm{sol}(X^a(t))$ is not an appropriate approximation to discuss the low energy physics of solitons and needs some extensions in $B\neq0$ case.
\subsection{Classification of zero modes}
\label{subsec:zero-classification}
In this section, we solve \eqref{eq:EOM_q} more strictly. First $\omega=0$ is a multiple root of $\Phi(\omega)$ in \eqref{eq:EOM-det}, since $\Phi(\omega)$ is a polynomial of $\omega^2$ as mentioned in Section \ref{subsec:fluctuation}.
According to the theory of ordinary differential equations, the solutions associated with $\omega = 0$ are some polynomials of $t$. So let us introduce the following polynomial ansatz
\begin{equation}
    q(t) = \sum_{k = 0}^{N}\frac{t^k}{k!}\eta_{(k)},
\end{equation}
where $N$ is finite and $\eta_{(k)}$ are arbitrary time-independent vectors. Substituting it into \eqref{eq:EOM_q}, we have
\begin{align}
\label{eq:EOM-q-polynomial}
    \sum_{k = 0}^{N-2}\frac{t^{k}}{k!}M\eta_{(k+2)}-\sum_{k = 0}^{N-1}\frac{t^{k}}{k!}B\eta_{(k+1)}+\sum_{k = 0}^N \frac{t^k}{k!}K\eta_{(k)} = 0.
\end{align}
For \eqref{eq:EOM-q-polynomial} to hold for arbitrary $t$, $\eta_{(k)}$ must satisfy
\begin{align}
    \begin{cases}
     K\eta_{(N)} = 0,\\
    B\eta_{(N)}-K\eta_{(N-1)} = 0,\\
    M\eta_{(k+2)}-B\eta_{(k+1)}+K\eta_{(k)} = 0,\quad k = 0,1,\cdots N-2.
    \end{cases}
\end{align}
If $N\geq2$, we have
\begin{equation}
    \eta_{(N)}^TM\eta_{(N)} -\eta_{(N)}^TB\eta_{(N-1)}+\eta_{(N)}^TK\eta_{(N-2)} = \eta_{(N)}^TM\eta_{(N)}+\eta_{(N-1)}^TK\eta_{(N-1)} = 0.
\end{equation}
Due to the positivity of $M$ and $K$, $\eta_{(N)}\equiv0$ for $N\geq2$. Then we have
\begin{align}
    K\eta_{(1)} & = 0\Rightarrow \eta_{(1)} \in\ker K,\\
    B\eta_{(1)} & = K\eta_{(0)}.
\end{align}
Therefore, the allowed polynomial solution is  
\begin{equation}
    q(t) = \sum_a[(X_0^a+V_0^a t)\xi_a+V_0^a\eta_a],
\end{equation}
where we replaced $\eta_
{(0)}$ and $\eta_{(1)}$ as
\begin{equation}
    \eta_{(0)}\to\sum_a(X_0^a\xi_a+V_0^a\eta_a),\,\eta_{(1)} \to \sum_aV_0^a\xi_a,
\end{equation}
and $\eta_a$ is taken to satisfy 
\begin{equation}
\label{eq:correction-vector}
    K\eta_a = B\xi_a.
\end{equation}
The $\eta_a$ term is known in many-body physics or chiral soliton models with vector mesons \cite{Cohen:1986va,Meissner:1987ge}. However, such a $\eta_a$ does not always exist for all $\xi_a$. Since it is obvious that $B\xi_a\in\Im K$, $\eta_a$ is well defined only for the zero modes $\xi_a$ satisfying 
\begin{equation}
\label{eq:dynamical-zero-condition}
    \xi^TB\xi_i = 0\,\,\mathrm{for\,\,arbitrary}\,\,\xi\in\ker K,
\end{equation}
These modes are called dynamical zero modes in \cite{Cherman:2005hy}. 

For zero modes that $B\xi_a$ are not orthogonal to $\ker K$, it turns out that the only polynomial solution of \eqref{eq:EOM_q} is a time-independent constant. Thus, these modes are named non-dynamical zero modes in \cite{Cherman:2005hy}. We take these modes orthogonal to dynamical zero modes as
\begin{equation}
\label{eq:dynamical-non-dynamical-orthogonal}
    \xi^TM\xi_i=0,
\end{equation}
for any dynamical zero modes $\xi_i$.

In this paper, we further divide non-dynamical zero modes into two groups. One is the space spanned by simultaneous eigenvectors of $M^{-1}B$ and $K$, denoted as $\qty{\xi_A}$ satisfying
\begin{equation}
\label{eq:cyclotron-modes}
    M^{-1}B\xi_A = -ib_A\xi_A,\,K\xi_A = 0,\quad b_A\in\mathbb{R}_{\neq0},
\end{equation}
and its orthogonal complement. For the former subspace, not only constant solutions but also time-dependent solutions of the form $e^{-ib_At}$ can appear. In this paper, we shall refer to these modes as cyclotron zero modes.
For the latter subspace, on the other hand, no time-dependent solution is allowed. As discussed in the next section, these modes induce constraints. Thus, we name them constraint zero modes.

In summary, there are two groups of zero modes:
\begin{itemize}
    \item Dynamical zero modes $\qty{\xi_i}$. We will use $i,j,\cdots\subset a,b,\cdots$ indices for the dynamical zero modes. 
    These modes satisfy $B\xi_i\in\Im K$, or equivalently \eqref{eq:dynamical-zero-condition}
     in order to possess the correlation vector $\eta_i$.
    The general solution of \eqref{eq:EOM_q} for these modes is
    \begin{equation}
    \label{eq:dynamical-solution}
        q(t) = \sum_i[(X_0^i+V_0^it)\xi_i+V^i_0\eta_i].
    \end{equation}
    Notice that $\eta_i$ satisfying \eqref{eq:correction-vector} for a given $\xi_i$ is not unique. One can add an arbitrary vector of $\ker K$ to $\eta_i$,
    \begin{equation}
    \label{eq:eta-redef}
    \eta_i\to\eta_i+\sum_{a}{\mathcal V_i}^a\xi_a.
    \end{equation}
    \item Non-dynamical zero modes. These zero modes do not satisfy $\eqref{eq:dynamical-zero-condition}$. Non-dynamical zero modes are separated into two subspaces based on the form of the solution of EOM \eqref{eq:EOM_q}.
    \begin{itemize}
        \item Cyclotron zero modes $\qty{\xi_A}$. The space of these modes is the subspace of non-dynamical zero modes spanned by the simultaneous eigenvectors of $M^{-1}B$ and $K$ as \eqref{eq:cyclotron-modes}. Note that $\xi_A$ are complex vectors and $\xi_A^*$ also belongs to cyclotron zero modes i.e. $(-b_A,\xi_A^*)$ also satisfy \eqref{eq:cyclotron-modes}. We label these modes with $A,B,\cdots$ indices to satisfy
        \begin{equation}
            (b_{-A},\xi_{-A})=(-b_A,\xi_A^*),
        \end{equation}
        for convenience.
        The solution of \eqref{eq:EOM_q} for these modes is
        \begin{equation}
        \label{eq:cyclotron-sol}
            q(t) = \sum_A\qty(\frac{1}{2}Z^A_0\qty(1+e^{-ib_At})+\frac{i}{b_A}W^A_0(e^{-ib_At}-1))\xi_A,
        \end{equation}
        where $Z^A$ and $W^A$ are complex constants that satisfy $(Z^{-A},W^{-A}) = ((Z^{A})^*,(W^A)^*)$
        \item Constraint zero modes $\qty{\xi_\alpha}$. $\alpha,\beta\cdots$ indices will be used for constraint zero modes.
    The space of these modes are defined as the orthogonal complement of cyclotron zero modes as 
    \begin{equation}
    \label{eq:orthogonal-constraint&cyclotron}
        \xi_\alpha^TB\xi_A = -ib_A\xi_\alpha^TM\xi_A = 0.
    \end{equation}
    The general solution of \eqref{eq:EOM_q} for these modes are time-independent
    \begin{equation}
    \label{eq:constraint-solution}
        q(t) = \sum_\alpha Y^\alpha_0\xi_\alpha.
    \end{equation}
    These modes lack dynamical degrees of freedom and instead generate constraints, as we will see later. 
    \end{itemize}
    As mentioned at \eqref{eq:dynamical-non-dynamical-orthogonal}, dynamical and non-dynamical zero modes are taken to be orthogonal. 
\begin{equation}
    \xi_i^TM\xi_A = \xi_i^TM\xi_\alpha = 0,
\end{equation}
for all $i$, $A$, and $\alpha$.
\end{itemize}

One can distinguish dynamical and non-dynamical zero modes from $B_{ab}$, the matrix element of $B$ on $\ker K$. According to \eqref{eq:dynamical-zero-condition} and \eqref{eq:orthogonal-constraint&cyclotron}, we have
\begin{equation}
\label{eq:B_K}
    B_{ab}=\xi_a^\dagger B\xi_b=\mqty(0 & 0 & 0\\0 & B_{AB}=-ib_A\delta_{AB} & 0 \\ 0 & 0 & \mathcal E_{\alpha\beta} := \xi_\alpha^TB\xi_\beta).
\end{equation}
Thus, dynamical and non-dynamical zero modes are associated with $\ker B_{ab}$ and $\Im B_{ab}$, respectively.
The separation between cyclotron and constraint zero modes needs to consider the whole $B$, which is far more difficult.

For remaining $\omega\neq0$ modes, we call them oscillating modes. Oscillating modes are related to soliton excitations. We label the oscillating modes with some non-zero indices $m,n,\cdots$ to satisfy
\begin{equation}
    (\omega_{-m},\xi_{-m}):=(-\omega_m,\xi_m^*),
\end{equation}
as well as cyclotron zero modes. These indices may take discrete integer values or represent continuous wave numbers, depending on the context.
We assume that there is no degeneracy for non-zero frequencies for simplicity. Then the oscillating mode part of the generic solution of \eqref{eq:EOM_q} is 
\begin{equation}
    q(t) = \sum_{m\neq0}\frac{\alpha_0^m}{\sqrt{2|\omega_m|}}e^{-i\omega_mt}\xi_m.
\end{equation}
Here, $\alpha_0^{m}$ are complex constants that satisfy $\alpha_0^{-m} = (\alpha_0^{m})^*$. 
Note that oscillating modes $\xi_m$ are not orthogonal to zero modes $\xi_a^TM\xi_m\neq0$ in general. 
\section{Collective coordinate quantization}
\label{sec:CCQ}
\subsection{Mode expansion and zero mode action}
In this section, we derive the effective zero mode action based on the full mode expansion.
As introduced in the previous section, the zero modes are divided into dynamical zero modes, cyclotron zero modes, and constraint zero modes when $B\neq0$, and they have different types of classical solutions. In addition, they appear differently in the action.

Including zero modes, we have the full general solution of \eqref{eq:EOM_q} and the momentum conjugate $p(t)$ as
\begin{align}
    \begin{split}
    \label{eq:EOM-sol-q}
    q(t) & = (X_0^i+V_0^it)\xi_i+V_0^i\eta_i+\sum_A\qty(\frac{Z_0^A}{2}(1+e^{-ib_At})+\frac{iW_0^A}{b_A}(e^{-ib_At}-1))\xi_A+Y_0^\alpha\xi_\alpha\\
     & +\sum_{m}\frac{\alpha^m_0}{\sqrt{2|\omega_m|}}e^{-i\omega_mt}\xi_m,
    \end{split}\\
    \label{eq:EOM-sol-p}
    p(t) & = \pdv{\dot{q}}L = M\dot{q}-\frac{1}{2}Bq \\
    & = V_0^iM\xi_i+\sum_A \qty(W_0^A-\frac{ib_A}{2}Z_0^A)e^{-ib_At}M\xi_A-\sum_{m}i\frac{\omega_m\alpha^m_0}{\sqrt{2|\omega_m|}}e^{-i\omega_mt}M\xi_m-\frac{1}{2}Bq.
\end{align}
Here, the summation over the zero mode indices $i$ and $\alpha$ is taken.
Now let us consider the following mode expansion analogous to \eqref{eq:EOM-sol-q} and \eqref{eq:EOM-sol-p}.
\begin{align}
    \label{eq:mode-expansion-q}
    q(t) & = X^i(t)\xi_i+V^i(t)\eta_i
    +\sum_AZ^A(t)\xi_A+Y^\alpha(t)\xi_\alpha+\sum_m\frac{\alpha^m(t)}{\sqrt{2|\omega_m|}}\xi_m,\\
    \label{eq:mode-expansion-p}
    p(t) & =V^i(t)M\xi_i+\sum_A\qty(W^A(t)-i\frac{b_A}{2}Z^A(t))M\xi_A-\sum_m\frac{i\omega_m\alpha^m(t)}{\sqrt{2|\omega_m|}}M\xi_m -\frac{1}{2}Bq(t).
\end{align}
Taking the inner products of \eqref{eq:EOM-sol-q}, \eqref{eq:EOM-sol-p} and $\xi_i$, $\xi_\alpha$, $\xi_m$, or $\eta_i$, we have
\begin{align}
\begin{split}
     g_{ij}X^j& +(\xi_i^TM\eta_j-\eta_i^TM\xi_j+\eta_i^TB\eta_j)V^j+\eta_i^TB\xi_\alpha Y^\alpha-\eta_i^TM\xi_A\qty(W^A+\frac{ib_A}{2}Z^A)
    \\ & = \xi_i^TMq-\eta_i^T\qty(p-\frac{1}{2}Bq),
    \end{split}\\
    g_{ij}V^j & = \xi_i^T\qty(p-\frac{1}{2}Bq),\\
    Z^A & +\frac{1}{\mu_A}\xi_A^\dagger M\eta_iV^i = \frac{1}{\mu_A}\xi_A^\dagger Mq\\
    W^A & +\frac{ib_A}{2\mu_A}\xi_A^\dagger M\eta_iV^i = \frac{1}{\mu_A}\xi_A^\dagger p\\
    -\mathcal E_{\alpha\beta}Y^\beta & -\xi_\alpha^TB\eta_iV^i = \xi_\alpha^T\qty(p-\frac{1}{2}Bq),\\
    \alpha^{m} & = \frac{\sqrt{2|\omega_m|}}{2\mu_m}\xi_m^\dagger\qty(Mq+\frac{i}{\omega_m}\qty(p-\frac{1}{2}Bq)).
\end{align}
Here $g_{ij} := M_{ij}+\eta_i^TK\eta_j$ and $\mathcal E_{\alpha\beta}=B_{\alpha\beta}$. $\mu_A$ and $\mu_m$ are the ``norm'' of $\xi_A$ and $\xi_m$ defined as
\begin{equation}
\label{eq:normalization}
    \mu_A:=\xi_A^\dagger M\xi_A,\,
    2\mu_m:= \xi_m^\dagger M\xi_m+\frac{\xi_m^\dagger K\xi_m}{\omega_m^2} = 2\xi_m^\dagger M\xi_m+\frac{1}{i\omega_m}\xi_m^\dagger B\xi_m,
\end{equation}
respectively. For the properties of the matrix elements of $M$, $B$, and $K$ used in the calculations, see the appendix \ref{appendix}.
Note that $g_{ij}$ and $\mathcal E_{\alpha\beta}$ are invertible matrices. This is easy to prove from the positivity of $M$ and $K$ and the definition of $\xi_\alpha$. As mentioned above, $\eta_i$ has the ambiguity shown in \eqref{eq:eta-redef}. Under this transformation, 
\begin{align}
    \xi_A^\dagger M\eta_i & \to \xi_A^\dagger M\eta_i+\mu_A{\mathcal V_{i}}^A\\
    \xi_\alpha^TB\eta_i- & \to \xi_\alpha^TB\eta_i+\mathcal E_{\alpha\beta}{\mathcal V_i}^\beta,\\
\begin{split}
    \eta_i^TB\eta_j+\xi_i^TM\eta_j-\xi_j^TM\eta_i & \to\eta_i^TB\eta_j + \xi_i^TM\eta_j-\xi_j^TM\eta_i+g_{ik}{\mathcal V_j}^k-g_{jk}{\mathcal V_i}^k\\
     & -ib_A\qty[{\mathcal V_j}^A\eta_i^TM\xi_A-{\mathcal V_i}^A\eta_j^TM\xi_A]\\
     & +\qty[{\mathcal V_j}^\alpha(\xi_i^TM\xi_\alpha+\eta_i^TB\xi_\alpha)+{\mathcal V_i}^\alpha(\xi_\alpha^TB\eta_j-\xi_j^TM\xi_\alpha)]\\
      & +{\mathcal V_i}^\alpha{\mathcal V_j}^\beta\mathcal E_{\alpha\beta}.
\end{split}
\end{align}
Thus, the terms on the left-hand sides can vanish by taking appropriate ${\mathcal V_i}^a$. As a result, one has
\begin{align}
    \label{eq:X}
    X^i & = g^{ij}\qty[\xi_j^TMq-\eta_j^T\qty(p-\frac{1}{2}Bq)],\\
    \label{eq:V}
    V^i & = g^{ij}\xi_j^T\qty(p-\frac{1}{2}Bq),\\
    \label{eq:Z}
    Z^A & = \frac{1}{\mu_A}\xi_A^\dagger Mq,\\
    \label{eq:W}
    W^A & = \frac{1}{\mu_A}\xi_A^\dagger p,\\
    \label{eq:Y}
    Y^\alpha & = -\mathcal E^{\alpha\beta}\xi_\beta^T\qty(p-\frac{1}{2}Bq),\\
    \label{eq:alpha}
    \alpha^{m} & = \frac{\sqrt{2|\omega_m|}}{2\mu_m}\xi_m^\dagger\qty(Mq+\frac{i}{\omega_m}\qty(p-\frac{1}{2}Bq)),
\end{align}
with $\eta_i$ satisfying
\begin{equation}
\label{eq:condition-eta}
    K\eta_i = B\xi_i,\,\eta_i^TM\xi_A = 0,\,\xi_\alpha^TB\eta_i = 0,\,\eta_i^TB\eta_j+\xi_i^TM\eta_j-\xi_j^TM\eta_i = 0.
\end{equation}
Here, $g^{ij}$ and $\mathcal E_{\alpha\beta}$ are the inverse matrix of $g_{ij}$ and $\mathcal E_{\alpha\beta}$, respectively.

Now let us expand the action. Consider the action in Hamiltonian formalism
\begin{equation}
\label{eq:path-integral-H}
    S = \int\dd t\qty(p^T\dot{q}-H(q,p)) = \int\dd t\qty[p^T\dot{q}-\frac{1}{2}\qty(\qty(p^T-\frac{1}{2}q^TB)M\qty(p+\frac{1}{2}Bq)+q^TKq)].
\end{equation}
Substituting the mode expansion \eqref{eq:mode-expansion-q} and \eqref{eq:mode-expansion-p} into \eqref{eq:path-integral-H}, we have
\begin{align}
    \begin{split}
        H & = \frac{1}{2}\qty[\qty(p^T-\frac{1}{2}q^TB)M^{-1}\qty(p+\frac{1}{2}Bq)+q^TKq]\\
        & = \frac{1}{2}g_{ij}V^iV^j+\sum_A\frac{\mu_A}{2}\qty(W^{-A}+\frac{ib_A}{2}Z^{-A})\qty(W^A-\frac{ib_A}{2}Z^A)\\
        & +\sum_{\omega_m>0}\frac{\mu_m\omega_m}{2}\qty(\alpha^m\alpha^{-m}+\alpha^{-m}\alpha^{m}),
    \end{split}\\
        p^T\dot{q} & = g_{ij}V^i\dot{X}^j+\sum_A\mu_AW^{-A}\dot{Z}^A+\frac{1}{2}\mathcal E_{\alpha\beta}Y^\alpha\dot{Y}^\beta+\sum_{\omega_m>0}\frac{i\mu_m}{2}(\alpha^{-m}\dot{\alpha}^m-\alpha^{m}\dot{\alpha}^{-m})+\dv{t}\qty(\cdots).
\end{align}
Here, $\cdots$ are some functions of $X^i$, $V^i$, $Y^\alpha$, and $\alpha^m$, and can be ignored. Thus, the action \eqref{eq:path-integral-H} can be decomposed into four sections,
\begin{equation}
\label{eq:action-expansion}
    \begin{split}
        & \int\dd t\qty(p^T\dot{q}-H(q,p)) = S[\alpha^m]+S[X^i,V^i]+S[Z^A,W^A]+S[Y^\alpha]\\
        S[\alpha^m] & = \sum_{\omega_m>0}\int\dd t\mu_m\qty[\frac{i}{2}\qty(\alpha^{-m}\dot{\alpha}^m-\alpha^m\dot{\alpha}^{-m})-\frac{\omega_m}{2}(\alpha^{-m}\alpha^m+\alpha^m\alpha^{-m})]\\
        S[X^i,V^i] & = \int\dd t\qty(g_{ij}V^i\dot{X}^j-\frac{1}{2}g_{ij}V^iV^j)\\
        S[Z^A,W^A]& = \sum_A\int\dd t\mu_A\qty[W^{-A}\dot{Z}^A-\frac{1}{2}\qty(W^{-A}+\frac{ib_A}{2}Z^{-A})\qty(W^A-\frac{ib_A}{2}Z^A)]\\
        S[Y^\alpha] & = \int\dd t\frac{1}{2}\mathcal E_{\alpha\beta}Y^\alpha\dot{Y}^\beta.
    \end{split}
\end{equation}
Each term is the action of oscillating modes, dynamical zero modes, cyclotron zero modes, and constraint zero modes, respectively. Note that there are no interactions between the oscillating modes and zero modes. The ``interaction'' term in \eqref{eq:eff-action-naive-with-nonzero} is due to a naive CCQ ansatz and does not mean that zero modes couple to other modes in the leading order.
$S[\alpha^m]$ is a sum of harmonic oscillator (HO) actions with frequency $\omega_m>0$ in terms of creation and annihilation operators $\alpha^m$. We can rewrite this term in a standard HO action by introducing 
\begin{equation}
    \alpha^m=:\frac{\sqrt{2|\omega_m|}}{2}\qty(x_{|m|}+\frac{i}{\omega_m}p_{|m|}).
\end{equation}
$S[X^i,V^i]$ is the action of the non-relativistic particle in the background metric $g_{ij}$ in the Hamiltonian formalism. In Lagrangian formalism, this term is  
\begin{equation}
    S[X^i] = \int\dd t\frac{1}{2}g_{ij}\dot{X}^i\dot{X}^j.
\end{equation}
$S[Z^A,W^A]$ is the sum of the action of the particle on a plane with a constant magnetic field in terms of the complex coordinates. The momentum conjugate of $Z^A$ is $\mu_AW^{-A}$, and we can rewrite the action as
\begin{equation}
    S[Z^A] = \sum_{b_A>0}\int\dd t\mu_A\qty[\dot{Z}^{-A}\dot{Z}^A-\frac{ib_A}{2}(Z^{-A}\dot{Z}^A-Z^{A}\dot{Z}^{-A})].
\end{equation}
$S[Y^\alpha]$ is quite different from the other terms. Since $Y^\alpha$ are absent in the Hamiltonian, 
this Lagrangian gives the constraint
\begin{equation}
    \phi_\alpha:=P_\alpha+ \frac{1}{2}\mathcal E_{\alpha\beta}Y^\beta =0,
\end{equation}
where $P_\alpha$ is the momentum conjugate of $Y^\alpha$. Calculating the Poisson brackets, we have
\begin{equation}
    \poissonbracket{\phi_\alpha}{\phi_\beta}_\mathrm{P}= \mathcal E_{\alpha\beta}.
\end{equation}
Since $\mathcal E_{\alpha\beta}$ is regular, $\phi_\alpha$ are second-class constraints and one can define the Dirac bracket as
\begin{equation}
    \poissonbracket{Y^\alpha}{Y^\beta}_\mathrm{D} = \mathcal E^{\alpha\beta}.
\end{equation}
As a result, the zero mode part of the action is 
\begin{equation}
\label{eq:zero-mode-Lagrangian}
\begin{split}
    S_\mathrm{zero} & := S[X^i]+S[Y^\alpha]+S[Z^A]\\
     & =\int\dd t\frac{1}{2}\qty[g_{ij}\dot{X}^i\dot{X}^j+\mathcal E_{\alpha\beta}Y^\alpha\dot{Y}^\beta+\sum_{b_A>0}\mu_A\qty(2\dot{Z}^{-A}\dot{Z}^A-ib_A(Z^{-A}\dot{Z}^A-Z^A\dot{Z}^{-A}))].
\end{split}
\end{equation}
Note that there are no interactions between the zero modes and oscillating modes.
The moduli metric $G_{ab}$ is obtained from the kinetic terms as
\begin{equation}
\label{eq:moduli-metric}
    G_{ab} = \mqty(g_{ij} & 0 & 0\\ 0 & \mu_A\delta_{AB} & 0 \\0 & 0 & 0).
\end{equation}
It is obvious that the moduli metric is not invertible.
\subsection{The generalization of the collective coordinate quantization}
\label{subsec:CCQ-method}
We now present the generalization of the CCQ. Comparing the action from the naive CCQ \eqref{eq:eff-action-naive} and \eqref{eq:zero-mode-Lagrangian}, the naive CCQ does not reproduce the moduli metric in two aspects. 
\begin{enumerate}[(i)]
    \item \label{enu:dynamical-prob}
    The dynamical zero mode component of the moduli metric is $g_{ij}$ in \eqref{eq:moduli-metric}. However, the evaluated metric from the naive CCQ is $M_{ij}$, which does not have the term $\eta_i^TK\eta_j$.
    \item \label{enu:constraint-prob}
    The naive CCQ does not distinguish zero modes and \eqref{eq:eff-action-naive} has the kinetic term of the constraint zero modes $\frac{1}{2}M_{\alpha\beta}\dot{Y}^\alpha\dot{Y}^\beta$, which actually does not exist in \eqref{eq:zero-mode-Lagrangian}.
\end{enumerate}
To solve the first problem \eqref{enu:dynamical-prob}, we add the $\eta_i$ term to the naive ansatz as
\begin{equation}
    q(t) =\sum_i(X^i(t)\xi_i+\dot{X}^i(t)\eta_i)+\sum_\alpha Y^\alpha(t)\xi_\alpha+\sum_AZ^A\xi_A.
\end{equation}
Substituting to \eqref{eq:action-fluctuation} and ignoring the time derivative term with order higher than three, we have
\begin{equation}
\begin{split}
    S = \int\dd t& \frac{1}{2}[M_{ij}\dot{X}^i\dot{X}^j+M_{\alpha\beta}\dot{Y}^\alpha\dot{Y}^\beta+\mathcal E_{\alpha\beta}Y^\alpha\dot{Y}^\beta\\
    & +\eta_i^TB\xi_j\dot{X}^i\dot{X}^j+\xi_i^TB\eta_j X^i\ddot{X}^j+\eta_i^TB\xi_\alpha\dot{X}^i\dot{Y}^\alpha+\xi_\alpha^TB\eta_i\ddot{X}^iY^\alpha
    -\eta_i^TK\eta_j\dot{X}^i\dot{X}^j]\\
     = \int\dd t& \frac{1}{2}[(M_{ij}+\eta_i^T(2B\xi_j-K\eta_j))\dot{X}^i\dot{X}^j+M_{\alpha\beta}\dot{Y}^\alpha\dot{Y}^\beta\\
     & +\mathcal E_{\alpha\beta}Y^\alpha\dot{Y}^\beta+2\eta_i^TB\xi_a\dot{X}^i\dot{Y}^\alpha]\\
      = \int\dd t & \frac{1}{2}\qty[g_{ij}\dot{X}^i\dot{X}^j+M_{\alpha\beta}\dot{Y}^\alpha\dot{Y}^\beta+\mathcal E_{\alpha\beta}Y^\alpha\dot{Y}^\beta].
\end{split}
\end{equation}
Here, we used \eqref{eq:condition-eta} and the cyclotron zero mode part is omitted. 

The second problem (\ref{enu:constraint-prob}) an be solved by removing the kinetic term of constraint zero modes $\frac{1}{2}M_{\alpha\beta}\dot{Y}^\alpha\dot{Y}^\beta$ by hand. 

In summary, the procedure of the generalized CCQ method is as follows.
\begin{enumerate}
    \item Find all the zero modes (or equivalently the collective coordinates) of the solutions.
    \item  Calculate $\xi_a^TB\xi_b$, $M\xi_a$, and $B\xi_a$ to classify the zero modes into dynamical zero modes, cyclotron zero modes, and constraint zero modes.
    \item For dynamical zero modes, find the $\eta_i$ that satisfy \eqref{eq:condition-eta}.
    \item Add the $\eta_i$ term to the naive CCQ ansatz $\phi(t) = \phi_\mathrm{sol}(X^a(t))$. The coefficient of the $\eta_i$ term is the first order of $\dot{X}^a$ and determined by substituting the ansatz into the EOM.
    \item Put the ansatz to the original action.
    \item Ignore the time derivative terms with order higher than three.
    \item Ignore the constraint zero mode components of the metric $M_{\alpha\beta}$.
    \item The effective action takes the form of
    \begin{equation}
        S = \int\dd t\qty(\frac{1}{2}\mathcal G_{ab}(X)\dot{X}^a\dot{X}^b+\mathcal A_{a}(X)\dot{X}^a).
    \end{equation}
    Note that $\mathcal G_{ab}(X)$ is not full-rank in general. Because of this, the effective action involves second-class constraints.
\end{enumerate}
In practical analysis, the most difficult part is the distinction between the cyclotron and the constraint zero modes. To determine the dynamical zero modes, only $B_{ab}$ is needed, and this is relatively easy to evaluate. The difference between the cyclotron and constraint zero modes, on the other hand, depends on whether the eigenvector of $M^{-1}B$ exists in $\ker K$ or not. Because of this, it is necessary to construct the exact $M$, $B$, and $\xi_a$ from the exact soliton configuration in general. 
\section{Application to the Skyrmion}
\label{sec:Skyrmion}
The $SU(3)$ Skyrme model with the WZW term is a good model for applying the generalized CCQ method discussed above. In this section, we use the following indices. 
\begin{equation}
\begin{split}
    \mathrm{4 dim\,\,spacetime\,\,indices}\,\,\mu,\nu,\cdots & = 0,1,2,3,\\
    \mathrm{5 dim\,\,spacetime\,\,indices}\,\,M,N,\cdots & = 0,1,2,3,4,\\
    SU(3)\,\,\mathrm{indices}\,\,a,b,\cdots & = 1,2,3,4,5,6,7,8,\\
    \mathrm{spatial\,\,and\,\,SU(2)\,\,indices}\,\,i,j,\cdots & = 1,2,3,\\
    SU(3)/(SU(2)\times U(1))\,\,\mathrm{indices}\,\,\alpha,\beta,\cdots & = 4,5,6,7.
\end{split}
\end{equation}
In addition, we use $\lambda_a$ as the Gell-Mann matrices.
\subsection{Skyrmion with WZW term and the naive CCQ}
First, we review the $SU(3)$ Skyrmion with the WZW term \cite{Witten:1983tw,Witten:1983tx,Mazur:1984yf} and the baryon spectrum from the naive CCQ \cite{Guadagnini:1983uv,Chemtob:1985ar,Diakonov:1997mm}.
The action of the model is 
\begin{align}
\label{eq:Skyrme-action}
    S[U] & = S_\mathrm{Skyrme}+S_\mathrm{WZW},\\
    S_\mathrm{Skyrme} & = \int\dd^4x\tr\qty[\frac{f_\pi^2}{16}L_\mu^2+\frac{1}{32e^2}\comm{L_\mu}{L_\nu}^2+\frac{f_\pi^2}{8}M(U+U^\dagger-2)],\\
    S_\mathrm{WZW} & = -i\frac{N_c}{240\pi^2}\int_{M_5}\dd^5x\epsilon^{KLMNO}\tr(L_K L_L L_M L_N L_O).
\end{align}
Here, $U(x)\in SU(3)$ is the pion field and $L_\mu:=\partial_\mu UU^\dagger$. $M_5$ is the 5-dimensional manifold with the boundary $\partial M_5$ being the 4-dim spacetime. $\epsilon^{KLMNO}$ is the Levi-Civita symbol with $\epsilon_{01234} = +1$. The parameters of the theory are the pion decay constant $f_\pi$, the Skyrme parameter $e$, the pion and quark masses $M$, and the number of colors $N_c$. We take the chiral limit $M \to0$ and the $ef_\pi/2 = 1$ unit. The EOM from \eqref{eq:Skyrme-action} is 
\begin{equation}
\label{eq:EOM-Skyrme}
    \frac{1}{e^2}\partial_\mu\qty(L^\mu-\frac{1}{4}\comm{L_\nu}{\comm{L^\nu}{L^\mu}})+i\frac{N_c}{24\pi^2}\epsilon^{\mu\nu\rho\sigma4}L_\mu L_\nu L_\rho L_\sigma = 0.
\end{equation}
The one-baryon configuration is written in the spherically symmetric ansatz 
\begin{equation}
\label{eq:B=1skyrmion}
    U(x) = U_\mathrm{sol}(x):=\mqty(\exp(if(r)n^i\lambda_i) & 0\\ 0 & 1),
\end{equation}
where $r := \sqrt{|x^i|^2}$ and $n^i:=x^i/r$. This ansatz is invariant under $SO(3)$ ``rotation''
\begin{equation}
\label{eq:skyrmion-ansatz-rotation}
    U(x) = V_RU(R^{-1}x)V^\dagger_R,V_R = \mqty(V_R^{SU(2)} & 0\\0 & 1)\,R\in SO(3).
\end{equation}
Here, $V^{SU(2)}_R$ is the $SU(2)$ representation of $R$.
$f(r)$ satisfies the EOM
\begin{equation}
\label{eq:B=1skyrmion-EOM}
(r^2+2\sin^2f)f''+2rf'
    +\sin2f\qty(f'^2-1-\frac{\sin^2f}{r^2}) = 0,\quad f'(r) = \dv{f(r)}{r},
\end{equation}
and $U_\mathrm{sol}(r\to\infty)\to1\Leftrightarrow f(\infty) = 0$.
The baryon number is 
\begin{equation}
\label{eq:Baryon-number}
\begin{split}
    N_B = \int\dd^3\vb*x\rho_{N_B}(\vb*x)& :=\frac{1}{24\pi^2}\epsilon^{0ijk4}\int\dd^3\vb*x\tr(L_iL_jL_k)\\
     & = -\frac{2}{\pi}\int_0^\infty\dd r\sin^2ff'
    = \frac{1}{\pi}f(0)-\frac{1}{2\pi}\sin2f(0).
\end{split}
\end{equation}
By requiring $N_B=1$, we obtain the boundary condition $f(0) = \pi$.

We define the fluctuation $u$ around $U_\mathrm{sol}$ as
\begin{equation}
    U(x) = (1+u(x))U_\mathrm{sol}(x)\Leftrightarrow u = UU_\mathrm{sol}^\dagger-1.
\end{equation}
The zero modes of the fluctuation are $u = L_i^\mathrm{sol}$ corresponding to the spatial translation, and $ u = \xi_{a} :=\frac{i}{2}\qty(\lambda_{a}-U_\mathrm{sol}\lambda_{a\neq8}U_\mathrm{sol}^\dagger)$ for $a\neq8$ corresponding to the $SU(3)_V/U(1)$ symmetry. Therefore, the moduli space of the $B=1$ Skyrmion is $\mathbb{R}^3\times SU(3)_V/U(1)$. We focus on the latter throughout this analysis. In the naive CCQ, we set the collective coordinate ansatz
\begin{equation}
\label{eq:naive-CCQ}
    U(x) = V(t)U_\mathrm{sol}(x^i)V^\dagger(t),\quad V(t) = \exp(iv^a(t)\frac{\lambda_a}{2})\in SU(3).
\end{equation}
Note that $V(t)$ has one gauge degree of freedom that does not change $U(x)$. 
\begin{equation}
\label{eq:gauge-transformation}
    V(t)\to V(t)\exp(-iv'_8(t)\frac{\lambda_8}{2}).
\end{equation}
Under this ansatz, the action \eqref{eq:Skyrme-action} reduces to
\begin{equation}
\label{eq:Skyrmion-eff-action-naive}
    S = S[U_\mathrm{sol}]+\int \dd tL_\mathrm{naive}[V] = S[U_\mathrm{sol}]+\int\dd t\qty[\frac{I}{2}(\Omega_R^i)^2+\frac{I'}{2}(\Omega_R^\alpha)^2-\frac{N_cN_B}{2\sqrt{3}}\Omega_R^8].
\end{equation}
Here, $I$ and $I'$ are
\begin{align}
    I = \int\dd^3\vb*x\rho_I(r) & := \frac{2}{3}\int\dd ^3\vb*x\frac{\sin^2f}{e^2}\qty(1+f'^2+\frac{\sin^2f}{r^2}),\\
    I' = \int\dd^3\vb*x\rho_{I'}(r) & := \int\dd^3\vb*x\frac{1}{2e^2}(1-\cos f)\qty(1+\frac{f'^2}{4}+\frac{\sin^2f}{2r^2}).
\end{align}
and $\Omega_R^a$ are defined as
\begin{equation}
    -iV^\dagger\dot{V} = \Omega_R^a\frac{\lambda_a}{2}.
\end{equation}
The momentum conjugate to $v^a$ are  
\begin{equation}
\label{eq:P_a}
    P_a = \pdv{L_\mathrm{naive}}{\dot{v}^a} = {e_a}^{i}I\Omega_i^R + {e_a}^\alpha I'{\Omega_\alpha}^8-\frac{N_c}{2\sqrt{3}}{e_a}^{8},\quad {e_a}^b(v):=\pdv{\Omega_R^b}{\dot{v}^a}.
\end{equation}
It is convenient to introduce $Q_a^R:={(e^{-1})_a}^{b}P_b$. These are the generators of the transformation
\begin{equation}
\label{eq:SU3_R}
    V\to VV_R^\dagger,\quad V_R\in SU(3)_R,
\end{equation}
because the Poisson brackets of $Q_a^R$s are 
\begin{align}
    \poissonbracket{Q_a^R}{V}_\mathrm{P} & = -{(e^{-1})_a}^{b}\partial_bV = -{(e^{-1})_a}^{b}V\qty(i{e_b}^c\frac{\lambda_c}{2}) = -iV\frac{\lambda_a}{2},\\
    \poissonbracket{Q_a^R}{Q_b^R}_\mathrm{P} & = f_{abc}Q_c^R. 
\end{align}
Here, we used the Maurer-Cartan identity
\begin{equation}
\label{eq:Maurer-Cartan}
\begin{split}
    & \partial_a(V^\dagger\partial_bV) -\partial_b(V^\dagger\partial_aV) = -V^\dagger\partial_aVV^\dagger\partial_bV+V^\dagger\partial_bVV^\dagger\partial_aV\\
    & \Rightarrow \partial_a{e_b}^c-\partial_b{e_a}^c = {e_a}^{a'}{e_b}^{b'}{f_{a'b'}}^c,\quad{e_b}^{b'}\partial_{b'}{(e^{-1})_a}^c-e_a^{a'}\partial_{a'}{(e^{-1})_b}^{c} = {f_{ab}}^{c'}{(e^{-1})_{c'}}^{c}.
\end{split}
\end{equation}
Thus, \eqref{eq:P_a} can be rewritten as
\begin{equation}
    Q_a^R=
    \begin{cases}
        I\Omega_R^i & \quad a = i\\
        I'\Omega_R^\alpha & \quad a = \alpha\\
        -\frac{N_c}{2\sqrt{3}} & \quad a = 8.
    \end{cases}
\end{equation}
The effective theory \eqref{eq:Skyrmion-eff-action-naive} has one constraint $\phi_8:= Q_8^R + \frac{N_c}{2\sqrt{3}} \equiv0$. This constraint reflects the gauge transformation \eqref{eq:gauge-transformation}. 

The effective action \eqref{eq:Skyrmion-eff-action-naive} has the following symmetry.
\begin{align}
    V(t) & \to V_LV(t),\quad V_L\in SU(3).\\
    V(t) & \to V(t)V_R^\dagger,\quad V_R = \mqty(V_R^{SU(2)} & 0\\0 & 1),\,V_R^{SU(2)}\in SU(2).
\end{align}
Note that the whole $SU(3)_R$ transformation \eqref{eq:SU3_R} is not a symmetry, and only the $SU(2)_R$ subgroup is preserved as a symmetry of this theory.

$V_L$ and $V_R$ transformations correspond to $SU(3)_V$ transformations and spatial $SO(3)$ rotations \eqref{eq:skyrmion-ansatz-rotation}, respectively.
The conserved charges of $SU(3)_V$ are
\begin{equation}
    Q^L_a :=-D_{ab}Q^R_b,\quad D_{ab}:=\frac{1}{2}\tr(V^\dagger\lambda_aV\lambda_b),
\end{equation}
and the conserved charges of $SO(3)$ rotation are $J_i:=Q_i^R$, which are equivalent to the spin $J_i$.
The Hamiltonian from \eqref{eq:Skyrmion-eff-action-naive} is 
\begin{equation}
    H = \frac{(Q_i^R)^2}{2I}+\frac{(Q_\alpha^R)^2}{2I'},
\end{equation}
and the energy spectrum for the representation of $SU(3)$ with Dynkin index $[p,q]$ is
\begin{equation}
    E = \frac{J(J+1)}{2I}+\frac{1}{2I'}\qty(\frac{1}{3}\qty(p^2+q^2+pq+3(p+q))-J(J+1)-\frac{N_c^2}{12}).
\end{equation}
Here, $J$ is the spin $(Q_i^R)^2 = J(J+1)$.

Physical states must satisfy the first-class constraint condition
\begin{equation}
\label{eq:1st-class-condtion}
    \phi_8\ket{\mathrm{phys}} = 0.
\end{equation}
This restricts the representation of $SU(3)$ to those which contain a state with right hypercharge $Y^R:=-\frac{2}{\sqrt{3}}Q_8^R = \frac{N_c}{3}$. For $N_c=3$, the allowed multiplets and their spin are 
\begin{equation}
    \begin{split}
        8 = & [1,1],\,J=\frac{1}{2},\\
        10 = & [3,0],\,J=\frac{3}{2},\\
        \overline{10} = & [0,3],\,J=\frac{1}{2},\\
        & \vdots
    \end{split}
\end{equation}
The octet represents the nucleon multiplet $N$. One of the decuplets with spin $\frac{3}{2}$ corresponds to the $\Delta$ baryons. The spin $\frac{1}{2}$ antidecuplet contains spin $\frac{1}{2}$, isospin $0$, and hypercharge $+2$ particle, which is called the $\theta^+$ pentaquark.
\subsection{The generalized CCQ}
Now let us adopt our CCQ method developed in section \ref{subsec:CCQ-method} to the Skyrme model. The EOM of the fluctuation $u$ is
\begin{equation}
\label{eq:EOM-q-Skyrme}
\begin{split}
    & -\frac{1}{e^2}\qty(\ddot{u}-\frac{1}{4}\comm{L^i_\mathrm{sol}}{\comm{L^i_\mathrm{sol}}{\ddot{u}}})+\frac{iN_c}{24\pi^2}\epsilon^{0ijk4}\anticommutator{\comm{\dot{u}}{L^i_\mathrm{sol}}}{L^j_\mathrm{sol}L^k_\mathrm{sol}}\\
    & +\frac{1}{e^2}\partial_i\qty(\partial^iu-\frac{1}{4}\comm{\partial_ju}{\comm{L^j_\mathrm{sol}}{L^i_\mathrm{sol}}}-\frac{1}{4}\comm{L_j^\mathrm{sol}}{\comm{\partial^ju}{L^i_\mathrm{sol}}}-\frac{1}{4}\comm{L_j^\mathrm{sol}}{\comm{L^j_\mathrm{sol}}{\partial^iu}})\\
    & +\frac{1}{e^2}\comm{\partial_iu}{L_\mathrm{sol}^i-\frac{1}{4}\comm{L_j^\mathrm{sol}}{\comm{L^j_\mathrm{sol}}{L^i_\mathrm{sol}}}}= 0,\quad L_\mathrm{sol}^i = \partial^iU_\mathrm{sol}U_\mathrm{sol}^\dagger,
\end{split}
\end{equation}
in the leading order. We can read off $M$, $B$ and $K$ in \eqref{eq:EOM_q} from this EOM, and their matrix elements are defined as
\begin{align}
\label{eq:M-skyrme}
    M(u,u') & :=-\frac{1}{e^2}\int\dd^3x\tr(uu'+\frac{1}{4}\comm{L^i_\mathrm{sol}}{u}\comm{L^i_\mathrm{sol}}{u'})\\
\label{eq:B-skyrme}
    B(u,u') := & -\frac{iN_c}{24\pi^2}\int\dd x^3\epsilon^{0ijk4}\tr(u\anticommutator{\comm{u'}{L^i_\mathrm{sol}}}{L^j_\mathrm{sol}L^k_\mathrm{sol}})\\
    K(u,u') := & -\frac{1}{e^2}\int\dd^3x\tr\Bigg(\partial_iu\partial^iu'\\
    & +\frac{1}{4}\qty(\comm{\partial_iu}{\partial_ju'}\comm{L_\mathrm{sol}^i}{L_\mathrm{sol}^j}+\comm{\partial_iu}{L^j_\mathrm{sol}}\comm{L^i_\mathrm{sol}}{\partial_ju'}+\comm{\partial_iu}{L_j^\mathrm{sol}}\comm{\partial^iu'}{L^j_\mathrm{sol}})\\
     & -\comm{u'}{\partial_iu}\qty(L^i_\mathrm{sol}-\frac{1}{4}\comm{L_j^\mathrm{sol}}{\comm{L^j_\mathrm{sol}}{L^i_\mathrm{sol}}})\Bigg)
\label{eq:K-skyrme}
\end{align}
Substituting $u 
 = \xi_a= \frac{i}{2}\qty(\lambda_a-U_\mathrm{sol}\lambda_aU_\mathrm{sol}^\dagger)$ into \eqref{eq:B-skyrme}, we have
\begin{equation}
    B_{ab} := B(\xi_a,\xi_b) = -\frac{N_c}{2\sqrt{3}}f_{ab8}
\end{equation}
The non-zero components of $f_{ab8}$ are $f_{458}=f_{678} = \frac{\sqrt{3}}{2}$. Therefore, $\qty{\xi_i}$ are dynamical zero modes and $\qty{\xi_\alpha}$ are non-dynamical zero modes.

Next we have to derive $\eta_i$  that satisfy \eqref{eq:condition-eta} for $\xi_i$ and distinguish whether $\qty{\xi_\alpha}$ are cyclotron or constraint zero modes. Putting $u 
 = iv^a(t)\xi_a$ into \eqref{eq:EOM-q-Skyrme}, we have
\begin{align}
    \frac{3}{2}\rho_I(\delta_{ij}-n_in_j)\ddot{v}^j & = 0,\\
    \label{eq:constraint-EOM-Skyrme}
    \rho_{I'}\ddot{v}^\alpha & + \frac{N_c}{2\sqrt{3}}\rho_{N_B}(1-\cos f)f_{\alpha\beta 8}\dot{v}^\beta = 0.
\end{align}
This is the explicit form of \eqref{eq:EOM-naive-zero} in this model.
The first equation indicates that $B\xi_i=0$ and it is obvious that $\eta_i\equiv0$. 
The second equation shows that $\qty{\xi_\alpha}$ are constraint zero modes. For \eqref{eq:constraint-EOM-Skyrme} to have time-dependent solutions, $\rho_I'\propto\rho_{N_B}(1-\cos f)$ is needed. However, this is not the case. This follows from the asymptotic behavior of $f(r)=\order{r^{-2}}$: $\rho_{I'}(r)=\order{r^{-4}}$ and $\rho_{N_B}(r) = \order{r^{-9}}$ in the chiral limit \cite{Skyrme:1962vh,Manton:1994ci}.
Therefore, the effective Lagrangian should not have the kinetic term proportional to $I'$. 
\begin{equation}
\label{eq:L_eff}
    S_\mathrm{eff} = \int\dd tL_{\mathrm{eff}} = \int\dd t\qty[\frac{I}{2}(\Omega_i^R)^2-\frac{N_c}{2\sqrt{3}}\Omega_8^R].
\end{equation}
The Hamiltonian from \eqref{eq:L_eff} and its eigenvalue are
\begin{equation}
    H_\mathrm{eff} = \frac{(Q_i^R)^2}{2I},\,E_\mathrm{eff} = \frac{J(J+1)}{2I}.
\end{equation}
Since $I'$ is lost in \eqref{eq:L_eff}, $Q_a^R$ are now 
\begin{equation}
    Q_a^R = 
    \begin{cases}
        I\Omega_R^i,& \quad a = i\\
        0,& \quad a = \alpha\\
        -\frac{N_c}{2\sqrt{3}}.& \quad a = 8
    \end{cases}
\end{equation}
This shows that the effective theory \eqref{eq:L_eff} has five constraints: $\phi_8=Q_8^R+\frac{N_c}{2\sqrt{3}}\equiv0$, which also appears from \eqref{eq:Skyrmion-eff-action-naive}, and $\phi_\alpha:=Q_\alpha^R\equiv0$, which are overlooked in the naive CCQ. The Poisson brackets between the constraints are 
\begin{equation}
    \mqty(
    \poissonbracket{\phi_\alpha}{\phi_\beta}_\mathrm{P} & \poissonbracket{\phi_\alpha}{\phi_8}_\mathrm{P}\\
    \poissonbracket{\phi_8}{\phi_\beta}_\mathrm{P} & 0
    )\approx\mqty({f_{\alpha\beta}}^iQ_i^R-{f_{\alpha\beta}}^8\frac{N_c}{2\sqrt{3}} & 0\\0 & 0),
\end{equation}
where $\approx$ is the weak equality. This implies that $\phi_8$ is first-class, while 4 $\phi_\alpha$s are second-class.

In the naive CCQ, $\phi_8=0$ is the only constraint and every state that satisfies \eqref{eq:1st-class-condtion} can be physical. On the other hand, however, there are also second-class constraints $\phi_\alpha$ in our method, and they give further restrictions.
To deal with the second-class constraints, a naive approach is to reduce the degrees of freedom by introducing the Dirac bracket. However, the $SU(3)$ symmetry becomes implicit, and the commutation rules become complicated. To deal with the second-class constraints in a covariant way, we adopt the Gupta–Bleuler quantization method, originally developed for quantum electrodynamics \cite{Gupta:1949rh,Bleuler:1950cy,Hasiewicz:1990xc}. 

First, let us introduce the following operators
\begin{equation}
    e_1^\pm:=\frac{1}{\sqrt{2}}(\phi_4\pm i\phi_5),\,e_2^\pm:=\frac{1}{\sqrt{2}}(\phi_6\pm i\phi_7).
\end{equation}
The second class constraints become $e_1^\pm = e_2^\pm = 0$.
The commutators between $e_1^\pm$ and $e_2^\pm$ are 
\begin{align}
    \comm{e_1^+}{e_1^-} & = \frac{1}{2}Q_3^R+\frac{\sqrt{3}}{2}Q_8^R,\\
    \comm{e_2^+}{e_2^-} & = -\frac{1}{2}Q_3^R+\frac{\sqrt{3}}{2}Q_8^R,\\
    \label{eq:e-comm}
    \comm{e_1^\pm}{e_2^\pm} & = 0,\\
    \comm{e_1^\pm}{e_2^\mp} & =  \pm\frac{1}{2}\qty(Q_1^R\pm iQ_2^R).
\end{align}
According to \eqref{eq:e-comm}, either a pair $(e_1^+,e_2^+)$ or $(e_1^-,e_2^-)$ can be regarded as the first class constraints, and the other as the gauge fixing condition. Then the physical state can be defined as
\begin{equation}
\label{eq:phys-condition}
     e_1^+\ket{\mathrm{phys}} = e_2^+\ket{\mathrm{phys}} = 0,\,\,\mathrm{or}\,\,e_1^-\ket{\mathrm{phys}} = e_2^-\ket{\mathrm{phys}} = 0.
\end{equation}
We understand the meaning of these conditions as follows. $e_1^\pm$ and $e_2^\pm$ are the operators that shift the weight $(Q_3^R,Q_8^R)$ by simple roots as
\begin{align}
\label{eq:weight-3_1}
    \comm{Q_3^R}{e_1^\pm} & =  \pm\frac{1}{2}e_1^\pm,\\
\label{eq:weight-3_2}
    \comm{Q_3^R}{e_2^\pm} & = \mp\frac{1}{2}e_2^\pm,\\
\label{eq:weight-8_1}
    \comm{Q_8^R}{e_1^\pm} & = \pm\frac{\sqrt{3}}{2}e_1^\pm,\\
\label{eq:weight-8_2}
    \comm{Q_8^R}{e_2^\pm} & = \pm\frac{\sqrt{3}}{2}e_2^\pm.
\end{align}
According to \eqref{eq:weight-8_1} and \eqref{eq:weight-8_2}, $(e_1^+,\,e_2^+)$ are the raising operators and $(e_1^-,e_2^-)$ are the lowering operators of $Q_8^R$, respectively. Hence \eqref{eq:phys-condition} means that the physical states are highest or lowest eigenstates of $Y^R\propto Q_8^R$ in the representation.

Combining with \eqref{eq:1st-class-condtion}, this indicates that $Y = \frac{N_c}{3}$ must be the maximum value of the hypercharge. In terms of Dynkin index $[p,q]$, this is expressed as
\begin{equation}
\label{eq:1st-2nd-condtion}
    p+2q = N_c.
\end{equation}
This condition is much stricter than \eqref{eq:1st-class-condtion}. For $N_c = 3$, only the nucleon octet $8 = [1,1]$ and the decuplet $10 = [3,0]$ related to $\Delta$ satisfy \eqref{eq:1st-2nd-condtion}. Therefore, all the other multiplets that appeared in the naive CCQ, including the anti-decuplet $\overline{10} = [0,3]$ with the $\theta^+$ pentaquark, are not allowed in our method. As a result, the infamous $\theta^+$ pentaquark is a mere artifact of the naive CCQ and never existed in the first place in the Skyrme model.

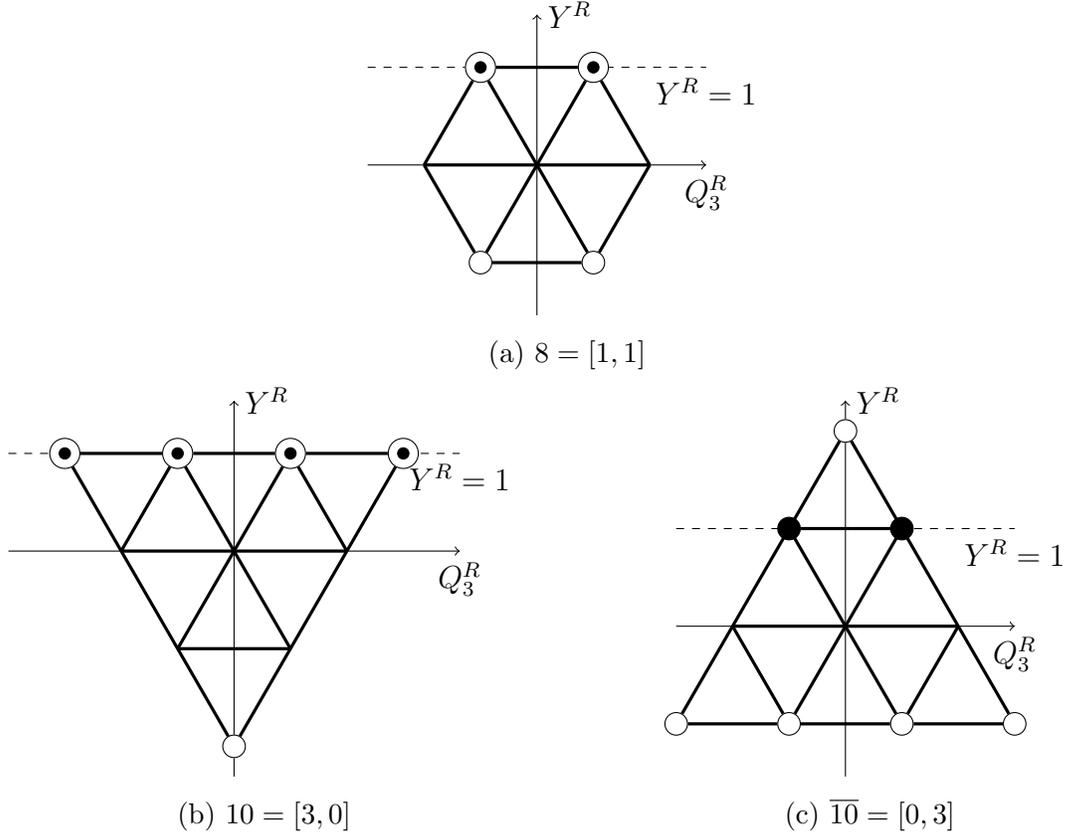
\begin{figure}[ht]
    \centering
    \begin{minipage}{0.55\columnwidth}
    \centering
    \begin{tikzpicture}
        \draw[->](0,0)--(9/4,0) node [below]{$Q_3^R$};
        \draw[->](0,-2)--(0,2) node [right]{$Y^R$};
        \draw[line width = 1.2pt](120:1.5)--(-60:1.5);
        \draw[line width = 1.2pt](-120:1.5)--(60:1.5);
        \draw[line width = 1.2pt](1.5,0)--(-1.5,0);
        \draw[line width = 1.2pt](60:1.5)--(120:1.5);
        \draw[line width = 1.2pt](-1.5,0)--(-120:1.5);
        \draw[line width = 1.2pt](-60:1.5)--(1.5,0);
        \draw[line width = 1.2pt](-120:1.5)--(-60:1.5);
        \draw[line width = 1.2pt](-1.5,0)--(120:1.5);
        \draw[line width = 1.2pt](1.5,0)--(60:1.5);
        \draw(0,0)--(-9/4,0);
        \draw[dashed](-9/4,3^1.5/4)--(9/4,3^1.5/4) node [below]{$Y^R=1$};
        \draw[fill,white] (60:1.5) circle[radius=0.2];
        \draw[fill,white] (120:1.5) circle[radius=0.2];
        \draw[fill,white] (-60:1.5) circle[radius=0.15];
        \draw[fill,white] (-120:1.5) circle[radius=0.15];
        \draw[fill] (60:1.5) circle[radius=0.075];
        \draw[fill] (120:1.5) circle[radius=0.075];
        \draw (60:1.5) circle[radius=0.2];
        \draw (120:1.5) circle[radius=0.2];
        \draw (-60:1.5) circle[radius=0.15];
        \draw (-120:1.5) circle[radius=0.15];
    \end{tikzpicture}
    \subcaption{$8=[1,1]$}
    \label{fig:diagram-8}
    \end{minipage}
    \begin{minipage}{0.45\columnwidth}
    \centering
    \begin{tikzpicture}
        \draw[->] (-3,0)--(3,0) node [below]{$Q_3^R$};
        \draw[->] (0,-3)--(0,2) node [right]{$Y^R$};
        \draw[line width = 1.2pt](120:1.5)--(-60:1.5);
        \draw[line width = 1.2pt](-120:1.5)--(60:1.5);
        \draw[line width = 1.2pt](1.5,0)--(-1.5,0);
        \draw[line width = 1.2pt](30:3^1.5/2)--(150:3^1.5/2);
        \draw[line width = 1.2pt](0,-3^1.5/2)--(150:3^1.5/2);
        \draw[line width = 1.2pt](0,-3^1.5/2)--(30:3^1.5/2);
        \draw[line width = 1.2pt](-120:1.5)--(-60:1.5);
        \draw[line width = 1.2pt](-1.5,0)--(120:1.5);
        \draw[line width = 1.2pt](1.5,0)--(60:1.5);
        \draw[dashed](-3,3^1.5/4)--(3,3^1.5/4) node [below]{$Y^R=1$};
        \draw[fill,white] (0,-3^1.5/2) circle[radius=0.15];
        \draw[fill,white] (60:1.5) circle[radius=0.2];
        \draw[fill,white] (120:1.5) circle[radius=0.2];
        \draw[fill,white] (30:3^1.5/2) circle[radius=0.2];
        \draw[fill,white] (150:3^1.5/2) circle[radius=0.2];
        \draw[fill] (60:1.5) circle[radius=0.075];
        \draw[fill] (120:1.5) circle[radius=0.075];
        \draw[fill] (30:3^1.5/2) circle[radius=0.075];
        \draw[fill] (150:3^1.5/2) circle[radius=0.075];
        \draw (0,-3^1.5/2) circle[radius=0.15];
        \draw (60:1.5) circle[radius=0.2];
        \draw (120:1.5) circle[radius=0.2];
        \draw (30:3^1.5/2) circle[radius=0.2];
        \draw (150:3^1.5/2) circle[radius=0.2];
    \end{tikzpicture}
    \subcaption{$10=[3,0]$}
    \label{fig:diagram-10}
    \end{minipage}
    \begin{minipage}{0.45\columnwidth}
    \centering
    \begin{tikzpicture}
        \draw[->] (-9/4,0)--(9/4,0) node [below]{$Q_3^R$};
        \draw[->] (0,-2)--(0,3) node [right]{$Y^R$};
        \draw[line width = 1.2pt](120:1.5)--(-60:1.5);
        \draw[line width = 1.2pt](-120:1.5)--(60:1.5);
        \draw[line width = 1.2pt](1.5,0)--(-1.5,0);
        \draw[line width = 1.2pt](-30:3^1.5/2)--(-150:3^1.5/2);
        \draw[line width = 1.2pt](0,3^1.5/2)--(-150:3^1.5/2);
        \draw[line width = 1.2pt](0,3^1.5/2)--(-30:3^1.5/2);
        \draw[line width = 1.2pt](120:1.5)--(60:1.5);
        \draw[line width = 1.2pt](-1.5,0)--(-120:1.5);
        \draw[line width = 1.2pt](1.5,0)--(-60:1.5);
        \draw[dashed](-9/4,3^1.5/4)--(9/4,3^1.5/4) node [below]{$Y^R=1$};
        \draw[fill] (60:1.5) circle[radius=0.15];
        \draw[fill] (120:1.5) circle[radius=0.15];
        \draw[fill,white] (0,3^1.5/2) circle[radius=0.15];
        \draw[fill,white] (-60:1.5) circle[radius=0.15];
        \draw[fill,white] (-120:1.5) circle[radius=0.15];
        \draw[fill,white] (-30:3^1.5/2) circle[radius=0.15];
        \draw[fill,white] (-150:3^1.5/2) circle[radius=0.15];
        \draw (0,3^1.5/2) circle[radius=0.15];
        \draw (-60:1.5) circle[radius=0.15];
        \draw (-120:1.5) circle[radius=0.15];
        \draw (-30:3^1.5/2) circle[radius=0.15];
        \draw (-150:3^1.5/2) circle[radius=0.15];
    \end{tikzpicture}
    \subcaption{$\overline{10}=[0,3]$}
    \label{fig:diagram-10bar}
    \end{minipage}
    \caption{The weight diagram of $8$, $10$, and $\overline{10}$ in a $SU(3)_R$ weight lattice. Black and white circles are the weights that satisfy \eqref{eq:1st-class-condtion} and  \eqref{eq:phys-condition}, respectively. Physical states that satisfies both \eqref{eq:1st-class-condtion} and \eqref{eq:phys-condition} are marked with $\odot$, which black and white circles match. \subref{fig:diagram-8} and \subref{fig:diagram-10} have 2 and 4 $\odot$s, respectively, while \subref{fig:diagram-10bar} do not. This means that the anti-decuplet $\overline{10}$ is not included in the physical state.}
    \label{fig:weight-diagram}
\end{figure}
\subsection{Comparison with other approaches}
In this section, we discuss the relation between our generalized CCQ and other methods : \cite{Cherman:2005hy}, bound state approach (BSA) \cite{Callan:1985hy,Callan:1987xt}, and rotation-vibration approach (RVA) \cite{Walliser:2005pi}.

The classification of zero modes into dynamical and non-dynamical zero modes was first presented in \cite{Cherman:2005hy} for the Skyrme-WZW model. According to this distinction, the upper bound of the hypercharge $Y\leq \frac{N_c}{3}$ is derived classically. The generalized CCQ is an extension of this by dividing non-dynamical zero modes further into cyclotron and constraint zero modes and removing the kinetic term of constraint zero modes. In our approach, the hypercharge bound is reproduced from the second class constraints quantum mechanically.

BSA and RVA are similar in ansatz, but give opposite results for the $\theta^+$ pentaquark. The ansatz of BSA and RVA are
\begin{align}
    U_\mathrm{BSA}(x) &= U_\mathrm{sol}(\vb*x)^\frac{1}{2}\exp(\frac{i}{f_\pi}\eta^a(x)\lambda_a)U_\mathrm{sol}^\frac{1}{2}(\vb*x),\\
    U_\mathrm{RVA}(x) & = V(t)U_\mathrm{sol}(\vb*x)^\frac{1}{2}\exp(\frac{i}{f_\pi}\tilde{\eta}^a(x)\lambda_a)U_\mathrm{sol}^\frac{1}{2}(\vb*x)V^\dagger(t), V(t)\in SU(3),
\end{align}
where $\eta_a$ and $\tilde{\eta}_a$ are introduced to describe meson excitations around the baryon ($\eta_a$ is denoted as $K_a$ in \cite{Callan:1985hy,Callan:1987xt}). $\tilde{\eta}_a$ in RVA is defined to be orthogonal to zero modes. The numerical analysis in BSA indicates that there are no bound states nor resonances corresponding to $\theta^+$. RVA, on the other hand, claims that $\theta^+$ resonance exists. The resonance structure is based on the phase shift $\delta$ with the ``background'' part $\bar{\delta}$ subtracted.   

To compare BSA and RVA with generalized CCQ, it is convenient to discuss them in terms of fluctuation $q$. For BSA, the EOM of $\eta^a$ and its mode expansion in \cite{Callan:1987xt} are actually equivalent to \eqref{eq:EOM_q} and \eqref{eq:EOM-eigenvector}, respectively. Hence, BSA approach treats the oscillating modes properly and is consistent with our generalized CCQ. As a result, there are no states that can be interpreted as $\theta^+$ in the whole spectrum.

In contrast,$\tilde{\eta}^a$ in RVA is associated with $q'$ in \eqref{eq:RVA-ansatz} with the orthogonal condition $\xi_a^TMq'=0$. 
This orthogonal condition causes several misunderstandings for the analysis in \cite{Walliser:2005pi}. 
First, $\tilde{\eta}^a$ does not represent the meson excitations completely. As mentioned in Section \ref{subsec:zero-classification}, the oscillating modes are not orthogonal to zero modes in general. Therefore, $\xi_a^TMq'=0$ is not a good assumption for oscillating modes, which are meson excitations in this case. Because of this, the meaning of the ``background'' phase shift $\bar{\delta}$ defined from these modes in \cite{Walliser:2005pi} is questionable, as pointed out in \cite{Cohen:2005bz}. Second, some of the ``interaction'' terms are fictitious. In RVA, there is an ``interaction'' term between the collective coordinates and $\tilde{\eta}^a$, which is treated perturbatively. However, this ``interaction'' term is an artifact due to the improper orthogonal condition. In addition, there are no reasonable explanations for the application of perturbation since the ``interaction'' term is $\order{N_c^0}$, which is of the same order as other terms of the Hamiltonian in large $N_c$. According to these misunderstandings, the discussion of RVA is not reliable compared with BSA or generalized CCQ.
\section{Conclusion}
\label{sec:conclusion}
We discussed the general collective coordinate quantization of solitons in the presence of topological terms. The basic idea of the method is the classification of zero modes. The moduli metric constructed from our method is non-invertible, and the effective action contains constraints in general. We applied our method to the Skyrmion with the WZW term and identified four additional constraints that were not accounted for in the naive approach. Using the Gupta-Bleuler quantization and group theoretical arguments, we demonstrated that only the nucleon and $\Delta$ baryon multiplets are permitted as physical states, which agrees with the classical analysis \cite{Cherman:2005hy}. In combination with BSA, we conclude that the $\theta^+$ state does not arise as a physical state in the Skyrme model, but rather is an artifact of the naive CCQ scheme.

Our analysis is based on the non-gauge systems and requires further developments to be applicable to solitons in gauge theories.
For gauge systems, the fluctuation around the soliton also has the gauge transformation and the mode expansion depends on the gauge fixing condition. Thus, we need to discuss the gauge invariance of the effective action derived in our method.
\section*{Acknowledgments}
I would like to thank Keito Shimizu, Hideo Suganuma, Shigeki Sugimoto, and Kei Tohme for valuable discussions.
This work was supported by JST BOOST, Grant Number JPMJBS2407. 
\appendix
\section{Inner products of the ``eigenvectors''}
\label{appendix}
We define the matrix elements of $M$ as
\begin{equation}
    M_{ab}:=\xi_a^\dagger M\xi_b,\,M_{am}:=\xi_a^\dagger M\xi_m,\,M_{ma}:=\xi_m^\dagger M\xi_a,\,M_{mn}:= \xi_m^\dagger M\xi_n.
\end{equation}
The matrix elements of $B$ and $K$ are defined as well. From the symmetry of $M$,$B$, and $K$, the matrix elements satisfy the following symmetry relations: 
\begin{align}
    \label{eq:symm-mel-M}
    M_{ab} & = M_{ba},\, M_{ma} = M_{a-m},\, M_{mn} = M_{-n\,-m},\\
    \label{eq:symm-mel-B}
    B_{ab} & = -B_{ba},\, B_{ma} = -B_{a-m},\, B_{mn} = -B_{-n\,-m},\\
    \label{eq:symm-mel-K}
    K_{ab} & = K_{am} = K_{ma} = 0,\,K_{mn} = K_{-n\, -m}.
\end{align}
Taking the inner product with \eqref{eq:EOM-eigenvector}, we have
\begin{align}
    -\omega_m^2M_{am}+i\omega_mB_{am} & = 0\Leftrightarrow B_{am} = -i\omega_mM_{am},\\
    -\omega_n^2M_{mn}+i\omega_nB_{mn}+K_{mn} & = 0.
\end{align}
When $m\neq n$, 
\begin{equation}
\label{eq:mel-relation-m=!n}
    \omega_n^2M_{mn}-i\omega_nB_{mn} = K_{mn} = K_{-n\,-m} = \omega_{-m}^2M_{-n\,-m}-i\omega_{-m}B_{-n\,-m} = \omega_m^2M_{mn}-i\omega_mB_{mn}.
\end{equation}
is satisfied from the symmetry \eqref{eq:symm-mel-M}, \eqref{eq:symm-mel-B}, and \eqref{eq:symm-mel-K}.
Therefore,
\begin{equation}
\label{eq:orthogonal-m=!n}
    K_{mn} = -\omega_m\omega_n M_{mn},\,B_{mn}=-i(\omega_m+\omega_n)M_{mn},
\end{equation}
for any $m\neq n$. For $m=n$, \eqref{eq:mel-relation-m=!n} is trivial and \eqref{eq:orthogonal-m=!n} is not valid because of the positivity. Instead, we can generalize \eqref{eq:orthogonal-m=!n} as
\begin{equation}
    K_{mn} = -\omega_m\omega_nM_{mn}+2\mu_m\omega_m\omega_n\delta_{mn},\, B_{mn} = -i(\omega_m+\omega_n)M_{mn}+2i\mu_m\omega_m\delta_{mn}.
\end{equation}
Here, $\mu_m$ is the ``norm'' of $\xi_m$ introduced in \eqref{eq:normalization}.

The cyclotron zero mode components are quite simple. First, it is easy to see that
\begin{equation}
    B_{AB} = -ib_BM_{AB} = -ib_AM_{AB}\Rightarrow M_{AB} = \mu_A\delta_{AB},
\end{equation}
with the norm $\mu_A$. From the condition \eqref{eq:dynamical-zero-condition}, we have
\begin{equation}
    0 = B_{iA} = -ib_AM_{iA}.
\end{equation}
For the inner products with the oscillating modes, we have
\begin{equation}
\label{eq:orthogonal-A,m}
    M_{Am} = \frac{i}{\omega_m}B_{Am} = \frac{b_A}{\omega_m}M_{Am}\Rightarrow M_{Am} = B_{Am} = 0.
\end{equation}
Here, we assume that $b_A\neq \omega_m$. In the case $b_A=\omega_m$, one can redefine $\xi_m$ to be orthogonal to $\xi_A$. Since the constraint zero modes and $\eta_i$ are defined to satisfy \eqref{eq:orthogonal-constraint&cyclotron} and \eqref{eq:condition-eta}, respectively, the cyclotron zero modes are orthogonal to both. Hence, they are orthogonal to all other modes.
\printbibliography[title = References]
\end{document}